\documentclass[5p,authoryear,times]{elsarticle}
\usepackage[english]{babel}
\usepackage{graphicx}
\graphicspath{{Figures/}}
\usepackage[breaklinks, colorlinks, citecolor=blue]{hyperref}
\usepackage{subfigure}
\usepackage{multicol,multirow}

\usepackage{amsmath,amssymb}

\usepackage{aas_macros}

\usepackage{xspace}
\usepackage{verbatim}
\usepackage{mathrsfs}
\providecommand{\spark}{{\tt Spark}\xspace}
\providecommand{\aspark}{\texttt{Apache} \texttt{Spark}\xspace}

\providecommand{\py}{{\tt python}\xspace}
\providecommand{\sca}{{\tt scala}\xspace}

\providecommand{\hp}{{\tt HEALPix}\xspace}

\providecommand{\colore}{{\tt CoLoRe}\xspace}
\newcommand{\citeg}[1]{\citep[e.g.,][]{#1}}
\def\q#1{\textquotedblleft{#1}\textquotedblright}
\newcommand{\SF}{\texttt{spark-fits}\xspace}

\defcitealias{SparkFITS:2018}{SparkFITS18}
\newcommand{\SFpap}{\citetalias{SparkFITS:2018}\xspace}
\newcommand{\tab}[1]{Table \ref{tab:#1}}

\usepackage{python-latex-highlighting/pythonhighlight}

\newcommand{\lcdm}{$\Lambda$CDM}

\providecommand{\text}[1]{\rm{#1}}

\newcommand{\begm}{\begin{pmatrix}}
\newcommand{\enm}{\end{pmatrix}}

\newcommand\ba{\begin{eqnarray}}
\newcommand\ea{\end{eqnarray}}
\newcommand\bea{\begin{eqnarray}}
\newcommand\eea{\end{eqnarray}}

\newcommand\be{\begin{equation}}
\newcommand\ee{\end{equation}}

\def\pmb#1{\setbox0=\hbox{#1}%
    \kern-.025em\copy0\kern-\wd0
    \kern.05em\copy0\kern-\wd0
    \kern-.025em\raise.0433em\box0}
\def\ltsima{$\; \buildrel < \over \sim \;$}
\def\gtsima{$\; \buildrel > \over \sim \;$}
\def\simlt{\lower.5ex\hbox{\ltsima}}
\def\simgt{\lower.5ex\hbox{\gtsima}}

\def\p2Y{\;_2Y}
\def\m2Y{\;_{-2}Y}

\newcommand{\ev}{\ensuremath{\mathrm{\,e\kern -0.1em V}}\xspace}

\def\eg{{e.g.}\xspace}

\newcommand{\sect}[1]{Sect.~\ref{#1}\xspace}

\usepackage{colortbl}
\definecolor{darkblue}{rgb}{0.0, 0.0, 0.55}
\usepackage{xcolor}
\usepackage[normalem]{ulem}

\begin{document}

%%%%%%%%%%%%%%%%%%%%%%%%%%%%%%%%%%%%%%
\begin{frontmatter}

% declarations for front matter
\title{Analysing billion-objects catalogue interactively: \\\aspark for
  physicists }

\author{S. Plaszczynski, J. Peloton, C. Arnault and J.E. Campagne}

\address{LAL, Univ. Paris-Sud, CNRS/IN2P3, Universit\'e Paris-Saclay, Orsay, France}

%\date{\today}

\begin{abstract}
\aspark is a Big Data framework for working on large
distributed datasets. Although widely used in the industry, it
remains rather limited in the academic community or often restricted to
software engineers. The goal of this paper is to show with
  practical uses-cases that the
technology is mature enough to be used without excessive programming
skills by astronomers or cosmologists in order to perform
standard analyses over large datasets, as those
originating from future galaxy surveys.\\
To demonstrate it, we start from a realistic simulation corresponding
to 10 years of LSST 
data taking (6 billions of galaxies). Then, we design, optimize and
benchmark a set of
\spark~\py algorithms in order 
to perform standard operations
as adding photometric redshift errors, measuring the selection function or
computing power spectra over tomographic bins. Most of the commands
execute  on the full 110 GB dataset within tens of seconds and can therefore be
performed interactively in order to design full-scale cosmological analyses.
A \texttt{jupyter} notebook summarizing the analysis is
available at \url{https://github.com/astrolabsoftware/1807.03078}.
\end{abstract}


\end{frontmatter}
%%%%%%%%%%%%%%%%%%%%%%%%%%%%%%%%%%%%%%

\section{Introduction}

%2 histoire spark
In 2002 \textsf{Google} released the \texttt{mapReduce} framework \citep[see e.g,][]{Dean:2008} 
a new parallel programming model exploiting efficiently many data
centre hardware. In 2006 its open source implementation 
 and many other tools related to the growing field of \q{Big data} emerged within the
\texttt{Hadoop} ecosystem \footnote{\url{https://hadoop.apache.org/}}.

In 2009 a research project started at UC. Berkeley
\citep{Zaharia:2012,Zaharia:2010} to overcome some
limitations met by \texttt{Hadoop}. In the next years, it captured many companies
attention due to order of magnitude better performances on huge distributed data
sets. It is today a very famous and active open source project named \spark owned by the
\texttt{Apache} foundation\footnote{\url{http://spark.apache.org/}}
and used by more than 1000 companies worldwide.

\spark was developed essentially in \sca, a multi-paradigm
  language that appeared in 2004 within the \texttt{java} ecosystem,  that revived \textit{functional programming} (FP)
\footnote{FP should not be compared to procedural or object-oriented
  programming which are both \textit{imperative} languages.}. FP deals with some old concepts
related to $\lambda-$ calculus where functions become the central
parts of programming.
This approach allows the \textit{lazy evaluation} mechanism which will be discussed 
in Sect.\ref{sec:spark}.
No knowledge of FP or \sca is necessary to use \spark.
Its API exposes efficient bindings to the \py or \texttt{R} languages.

While under the influence of
business and web companies, Big Data technologies were emerging,
the data trend in astronomy was also rising rapidly. Typical
surveys today reach Terabytes of raw data and Petabytes
in a very near future \citeg{Zhang:2015a}.
On the cosmological side, even after the reduction of images to user
catalogues the \texttt{BOSS} experiment
\footnote{\url{http://www.sdss3.org/index.php}} already imaged
millions of 
galaxies. While the next generation of spectroscopic survey
\texttt{DESI}\footnote{\url{http://desi.lbl.gov}} 
is planned to image tens of millions of galaxies, the \texttt{DES} photometric
survey\footnote{\url{https://www.darkenergysurvey.org}} already provides such a data volume.
In the next decade \texttt{LSST}\footnote{{\url{https://www.lsst.org}}} should reveal the properties
of billions of galaxies that will be collected over 10 years.
Moreover the study of the galaxy distribution properties requires tens
to hundred times more simulated samples (mock catalogs).

We have also entered the Big Data area in astronomy not only for image
treatment but also for analysing the output catalogues, which will
be more and more limited by I/O throughput. Fortunately, such
amounts of data are \textit{not} impressive for today \spark industry
standards.

Some exploratory work on using \spark in astronomy has already been shown by experts
who ingested some large astronomical catalogues and organized them in a way where
very optimized queries (as catalogues cross-match) can be performed
efficiently \citep{zecevic:2018,BrahemYZ18,Carretero:2017,2017PASP..129b4005R}. Although very promising for
the future, this database-driven approach shows only one aspect of \spark possibilities.

% ref spark en science.

\begin{figure}
  \centering
  \includegraphics[width=.4\textwidth]{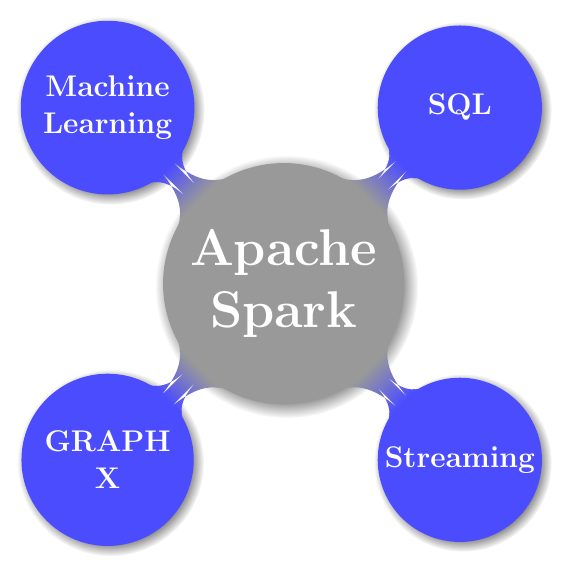}
  \caption{Apache Spark provides four different modules. The SQL
    module focuses on the manipulating structured data such as tables.
    The Machine Learning module provides tools to perform distributed
    machine learning. The Streaming module allows to write streaming
    application. The GraphX module is the part of the API for graphs
    and graph-parallel computation. In this paper, we focus only on
    the Apache Spark SQL module.} 
  \label{fig:spark_building_blocks}
\end{figure}

Indeed \spark is a large \textit{framework} and Fig.
\ref{fig:spark_building_blocks} shows its core modules, which can be complemented by numerous libraries. 
For this paper we will focus essentially on the
\texttt{SQL} part, allowing users to perform efficient queries and analyses
on catalogs. 
But \spark also offers machine-learning possibilities (\texttt{MLLIB}
module \citet{Meng:2015}) which can be complemented by external
libraries\footnote{see for instance \url{https://maxpumperla.com/elephas}.}.
A first application of the graph analysis module (\texttt{Graph-X})
to investigate the topology of cosmological simulations was
recently presented in \cite{Hong:2019}, while \textit{Streaming} possibilities
open exciting perspectives for the real-time treatment of alerts by
brokers in astronomy
\footnote{as in \url{https://fink-broker.readthedocs.io/en/latest/}}. 
An introduction
on using these different modules can be found in \citet{ZecevicBook}.

\aspark was also used to
perform distributed processing of astronomical images \citep{Zhang:2015,Wiley:2011} or
to produce mock catalogues from huge N-body simulations
\citep{Carretero:2017}. This illustrates the potential of the framework
when used for fundamental sciences.

Our goal here is not to report the impressive results obtained by
Spark experts, but to address the needs of end-users (physicists) who
have little or no knowledge about Spark. 
We will not elaborate on the theory that
  can be found in many articles or textbooks \citep[a classical
  introduction being][]{LearningSpark:2015}
but try to show with concrete examples how this framework offers a \textit{simple} way to
achieve excellent results for analysing today's and upcoming catalogs
in astronomy on computing clusters. Today's main language for astronomical data
analysis being \py, we will focus on that language. We will use some familiar
concepts as dataframes and emphasise how external \py codes can be
reused. The use-case we develop is the following: one obtains a (large) set
of astronomical files, not necessarily produced by the user,
and would like to have a look at some of the variables and perform some
simple analysis over the entire dataset.
As an example, we choose to use a catalogue of 6 billion of
galaxies generated with a fast simulation and show how to perform
easily an interactive tomographic analysis in a very reasonable time.
Since a very common format in astronomy is FITS, we first developed
an efficient connector named \SF
\footnote{\url{https://astrolabsoftware.github.io/spark-fits}} 
described in \citet{SparkFITS:2018}, hereafter called \SFpap.

After presenting in a physicist's language what \spark is
and what its advantages are for data analysis in \sect{sec:spark}, we
describe how to start using \spark in \py in \sect{sec:tools}.
Then we design and optimize several \spark commands in
\sect{sec:ana} with increasing complexity leading to a
full tomographic analysis over a simulated set
of 10-years of LSST-like galactic data (\sect{sec:tomo}). Several options are
explored and performances discussed. \ref{sec:appendix} presents a
 more realistic treatment of photometric uncertainties.

\section{What is \spark and why to use it}
\label{sec:spark}

The main trend for analysing large astronomical datasets in the last
two decades was to use High Performance Computing (HPC) approaches on
super-computers.
The challenge in using such a framework consists in
optimising the  \textit{arithmetic efficiency} (number of data moves
over number of operations) on more and more complex architectures and
requires a high degree of skills. This can be performed with 
the Message-Passing-Interface (MPI, \citet{MPI:1996}) but requires a careful tuning of
the code for data exchanges between the nodes.

Starting with \texttt{mapReduce}, a new approach to address these problems has
focused on High Through- put Computing (HTC) in data centres.
Accent is put 
%on data locality, i.e
interacting with nodes that are \q{close} to the data and minimizing the 
traffic between them. 
A key requirement for this is robustness and reliability, which is now
achieved by performing computation in a fault-tolerant way on hardware
which may not be fault-tolerant. 
A more in-depth overview can be found in \citet{Stickey:2015}.

\spark is now for many years the leading solution in this domain.
Hosted by the \texttt{Apache foundation}, the code is
open-source and its community very active. 
In very general terms, \q{\aspark is a cluster computing platform
designed to be fast and general}\citep{LearningSpark:2015}. 
Although available for personal usage on laptops, it will reveal its
full potential when running on a cluster on some 
large amount of data \footnote{by large we mean beyond several tens of
  Gigabytes.} within a distributed file system.

To introduce the necessary vocabulary, we outline some of the features
of the \spark solution and refer the reader to
\citet{LearningSpark:2015,Tang:2018,SparkFITS:2018} for a more
thorough technical introduction. 

In \spark, the user program, called the \texttt{driver}, launches multiple \texttt{workers}. Each worker can
 hold several \texttt{executors} (processes launched for an
 application) that execute a number of \texttt{tasks}.\footnote{In the
examples below, we restrict the case to one executor per worker,
with several tasks.} Workers read data by blocks (possibly from
  a distributed file system) and can persist computed objects in
  memory. In a typical work flow the \spark scheduler sends a set of
  tasks to the workers in order to compute locally over the nodes holding the
  data or in memory if they have been cached. The
  reduced outputs are then returned to the driver that can
  perform some final combination.

Although quite an evolved process, on a practical level the user
only needs to interact with \texttt{RDD}s (Resilient
Distributed Datasets, 
\citet{Zaharia:2012})
which are abstractions representing a distributed collection of
objects. They provide a low level access to \spark features.
Scientists will typically prefer to work with \texttt{DataFrames} (from the SQL module),
which are built on top of RDD but with data organized into named
columns and which also include the benefits of the Spark SQL's execution engine
(Sect. \ref{sec:df}).

Using Spark presents a number of advantages that are particularly well
suited to data analysis tasks in the physical sciences, as we discuss
below.

\paragraph{Simple parallelisation}
The use of the \texttt{mapReduce} paradigm allows a coarse level of parallelisation without ever writing
complicated code or directives.
This is best illustrated by an example. Suppose you are looking for the maximum
value of a huge set of numbers. The driver requires workers
to compute their maximum value. The results from this \texttt{map} step
are then returned and combined in the \texttt{reduce} step, here by taking
the maximum of their values.
Such a behaviour could also be implemented with other mechanisms
such as \texttt{MPI}, but would require writing
some extra piece of code while, quite remarkably, it is \textit{builtin}
feature in \spark.

\paragraph{Optimization}
In \textit{imperative} languages like  for example \texttt{C},
  \texttt{C++} or 
\texttt{Fortran}, each instruction generates some low level commands
that are immediately executed.
\spark uses the mechanism of \textit{lazy evaluation}, meaning that
most instructions (called \textit{transformations}) do \textit{not}
immediately trigger execution, but instead updates the construction of
a Direct Acyclic Graph or DAG which in physicists language represents a
pipeline.  
It is often a common surprise to newcomers to see how fast some Spark
commands seem to perform, but this is only because the pipeline has
not yet been closed by specifying an action, so nothing was actually
run (see an example in Sect. \ref{sec:load}).
Then, the \py front-end to \spark can be seen as a
way of declaratively specifying a calculation that is then passed to
the \spark layer for evaluation. 
An interesting aspect of this is that knowledge of the full pipeline
structure allows \spark to perform some precise optimizations based
on a set of logical/physical rules before running it.  
As long as one makes use of \spark functions, the user is relieved of
the burden of code optimisation such as accessing different parts of
data in different places.

\paragraph{Cache and interactive analysis}
Maybe the most interesting feature for the end-user analysis is the ability of
putting the data in cache. While quite a complex operation
behind the hood, on
the user side this is as simple as calling the \texttt{cache()}
function.
Even when the cluster does not allow all the data to be held in
memory, putting some part in cache and spilling the rest over disk can
still be interesting (see Fig. 3 of \SFpap). 
A key point is that in general, you do not need to put all the
variables in cache, only those that will be accessed several times.
The memory requirements are
then drastically reduced.

Once the relevant variables are cached, all computations can be
carried out efficiently enough that an interactive analysis becomes
generally possible. 
For demonstration purposes in the following sections, we have put 110 GB of data in memory on a small
cluster, but one can achieve easily 1 TB on large centres like the 
National Energy Research Scientific Computing Center
(NERSC)\footnote{\url{https://www.nersc.gov}}.

\paragraph{Performances and scaling}
\spark, as a largely used industrial standard, gives generally excellent
performances for analysing large datasets.
We do not claim that \spark will always give the ultimate performance as depending on the 
specific domain some implementations with \texttt{MPI} or using relational databases,
carefully tuned by experts, may outperform it. But having at disposal
such a simple tool to reach such high level of efficiency is a great
opportunity for physicists facing Big Data challenges.

The attractiveness of Spark also resides in the fact that one can
design some analysis of a partial set of data on a personal laptop and
later port it to large data centres knowing that in most cases
performance will scale.

\section{Working with \spark}
\label{sec:tools}

As we saw previously, \spark's basic abstraction is the RDD
which represents a coarse-grained distributed collection of objects
which allow in-memory fault-tolerant computations.
\textit{Dataframes} were later introduced in 2015 
within the \texttt{SQL} module \citep{SparkSQL:2015}. 
They add to the RDD the knowledge of its data structure which allows a higher level of
optimization. Although one can retrieve an RDD from a dataframe, one should
always try to use the latter in the analyses, especially for \py applications. 
Indeed the RDD performances are much worse in \py than in
the native \sca language. This is not any more the
case when working with dataframes, for which similar performances are
obtained in most cases as shown in \sect{sec:perf}.

Most of astronomers use today the \py language 
for their analyses and we emphasize that
\texttt{pyspark} can be run within \texttt{ipython} or a
\texttt{jupyter notebook}. We will then essentially discuss the \py interface
although some comparison with \sca performances will be later shown.
In our opinion, the scala language remains an interesting option to be
looked at and we highlight some of the reasons why it may be useful to
astronomers below: 
\begin{itemize}
\item there exist some \spark kernels in \sca to work in \texttt{jupyter}
  notebooks\footnote{\eg \url{https://almond.sh/docs/usage-spark}},
\item the \hp package \footnote{\url{http://healpix.sourceforge.io}}
 can be build in
\texttt{java} (and therefore used in \sca) although not including all 
functionalities, 
\item the \texttt{JEP} package \footnote{\url{https://pypi.org/project/jep}}
  allows to make calls to external \py modules as \texttt{numpy} or 
  \texttt{matplotlib}, and the library \texttt{Py4J}\footnote{\url{https://www.py4j.org/}} provides an efficient bridge between \py and \sca.
\end{itemize}

Concerning data formats, \SF allows the reading of FITS distributed
binary tables and also in recent versions of images.
Some connectors also exist to read the \texttt{HDF5}
format (see references in \SFpap).

\section{A full interactive analysis}
\label{sec:ana}

We present a suite of rather standard operations \footnote{A \texttt{jupyter notebook} is available at \url{https://github.com/astrolabsoftware/1807.03078}}
an end-user can perform
using the output of a 10 years LSST simulation presented in \sect{sec:colore}. 
Commands were run interactively in the \texttt{pyspark} shell
on a cluster described in \sect{sec:infra}.
They are explicitly shown as boxed inline code with their output following right after.
We benchmark different options in some cases and focus more on performances
in \sect{sec:perf}.

\subsection{Simulation}
\label{sec:colore}

\begin{figure}
  \centering
  \includegraphics[width=\linewidth]{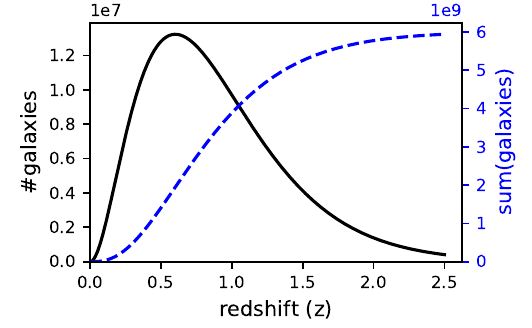}
  \caption{Density number of galaxies (black, left axis) and their cumulative sum
    (dashed-blue, rights axis) generated by
    \colore corresponding to 10 years of LSST data-taking.} 
  \label{fig:cumgals}
\end{figure}

In order to work with physics-oriented data, we built a
catalogue of galaxies using the \colore fast simulation 
\footnote{\url{https://github.com/damonge/CoLoRe}} corresponding to 10 years of LSST data-taking. 
Point-like galaxies are generated in the $z\in [0,2.5]$ redshift range
assuming a standard \lcdm\ cosmology, with a selection function coming
from the 
LSST/DESC 2pt\_validation working group
\footnote{\url{https://github.com/LSSTDESC/2pt_validation}} and shown on
Fig.\ref{fig:cumgals}. 

The generated  catalogue consists of 6 billions of galaxies
with their types, RA/DEC positions, cosmological redshifts
and the redshift-space distortion (RSD) displacements. 
The \colore simulation was run at \texttt{NERSC} and produced 32 FITS files of 112 GB in
total that were imported to our \texttt{HDFS} cluster\citep{HDFS:2010}. 
By importing, we mean simply copied to the HDFS cluster without any post-processing.
Then, \SF gives transparent access to all the \spark advantages.

This procedure is mainly pedagogical since it introduces latencies for transferring data to another cluster.

Whenever possible, one should instead attempt to produce and analyse
the data on the same cluster, which would have been possible on the
NERSC supercomputer since it offers a Spark API (more in \SFpap).

\subsection{Infrastructure}
\label{sec:infra}

The cluster we used for this work, located at Universit\'e Paris-Sud in
France
\footnote{\url{https://www.informatique-scientifique.u-psud.fr/services/spark.html}}
is rather modest in order to illustrate the fact that you do
not need huge resources to achieve spectacular results. 
It consists of nine 36 GB machines, each with 18 cores, running over
\texttt{HDFS}.
Since the total memory fraction dedicated
to the cache is set to 0.6, the usable memory amount ( $\simeq$ 170 GB) is
largely sufficient to hold our full dataset.
Then, we used the following set-up
\begin{itemize}
\item 1 driver (4 GB RAM)
\item 8 executors (ie. workers) each using 17 cores and 30 GB of RAM.
\end{itemize}
The \py interactive shell is run as followed
\begin{verbatim}
pyspark --driver-memory 4g \
        --total-executor-cores 136 \
        --executor-cores 17 \
        --executor-memory 30g 
\end{verbatim}

\subsection{ \spark analysis}

\subsubsection{Using dataframes}
\label{sec:df}
Dataframes in \spark come from the \texttt{Spark.SQL} module.
Similar to the \texttt{pandas} one
\footnote{\url{https://pandas.pydata.org/}}, they can be viewed as a set of named \textit{columns} over which you
can perform operations, but in a distributed environment.
Some native \spark functions act
on them and should be used as much as possible since they have been
very optimized. They are available through the following \verb= import= command
\begin{python}
from pyspark.sql import functions as F
\end{python}

Once substantial data reduction has been achieved, we can 
recover a standard \texttt{pandas} dataframe with the \verb=toPandas()=
method, which opens the door to further standard \py analysis. It is worth mentioning that \verb=toPandas()=
returns to the driver memory, so one should keep in mind that the data reduction (\verb=reduce= methods) 
must be performed before.

\subsubsection{Reading the data}
\label{sec:load}

We begin by loading the (set of) FITS files using \SF. \colore FITS format
stores separately the cosmological ( \verb=Z_COSMO=) 
and RSD (\verb=DZ_RSD=)
redshifts, but since we only want to work on their sum, we
construct the \verb=z= column on the fly:
\begin{python}
gal=spark.read.format("fits")\
    .option("hdu",1)\
    .load("hdfs:path/to/fits/directory")\
    .select("RA","Dec", \
    (F.col("Z_COSMO")+F.col("DZ_RSD"))\
    .alias("z"))
\end{python}
This represents a (\spark) dataframe object.
We show two ways of accessing columns in the \texttt{select}
function, either simply through their names (strings) or, when some
operations are to be performed, through \verb=F.col(..)= that returns
\texttt{Columns} objects. Other dataframe standard ways to access columns are
through \verb=gal.RA= or \verb=gal['RA']=.
Note the use of the \q{z} alias to rename the
new column. 

You can now print the dataframe schema:
\begin{python}
gal.printSchema()
\end{python}
\begin{verbatim}
root
 |-- RA: float (nullable = true)
 |-- Dec: float (nullable = true)
 |-- z: float (nullable = true)
\end{verbatim}

Following the principle of lazy evaluation, it is important to realize
that, at this level, data is not (yet) physically
loaded: only the FITS header is read and the DAG updated.

\subsubsection{Adding photometric smearing}
\label{sec:PZ}

We would like to add now the effect of the photo-z (PZ) resolution of
the instrument. For \texttt{LSST} this will be close to a Gaussian
and the upper requirement for Large Scale Structure analysis is given
in \citet{SRD:2018} as $\sigma_z=0.03(1+z)$ .
A lot of work based on template-fitting or Machine Learning
techniques shows that the actual distribution is more
complicated than a simple Gaussian. 
However to not distract the reader from the main goal of this paper,
we have relegated to \ref{sec:appendix} a more realistic discussion of this
effect and how it can be implemented within \spark.

For the Gaussian case, we simply add a column to the \texttt{gal} dataframe according to

\begin{python}
from pyspark.sql.functions import randn
gal=gal.withColumn("zrec",
    gal.z+0.03*(1+gal.z)*randn())
    .astype('float'))
\end{python}

Again, when this last command is executed the DAG is further filled
but no data has been physically generated.

Let us now show 5 samples, which triggers an action:
\begin{python}
gal.show(5)
\end{python}
\begin{verbatim}
\end{verbatim}
\begin{verbatim}
+---------+---------+---------+---------+
|       RA|      Dec|        z|     zrec|
+---------+---------+---------+---------+
|225.80168|18.519966|2.4199903| 2.414322|
|225.73839|18.588171|2.4056022|2.2913096|
|225.79999|18.635067| 2.396816|2.3597262|
|225.49783|18.570776|2.4139786|2.3434482|
|225.57983|18.638515|2.3995044|2.3826954|
+---------+---------+---------+---------+
only showing top 5 rows
\end{verbatim}

This happens within seconds (see \tab{perf}). How is that possible?
This is the very idea of lazy evaluation: if you only want to look at
a few samples is it worth loading all (110 GB) of the data? Here \spark
analyses the full pipeline, optimizes it and only physically read the
very first block

\subsubsection{Caching}
\label{sec:cache}

Now that we have defined which data we want to use in our analysis, we
put them in cache. 
This is achieved with the \verb=cache()= function. Some finer level of
details can be obtained with the \verb=persist(level)= function that
allows to specify the storage level. \verb=cache()= corresponds to
\verb=persist(MEMORY_ONLY)=. You may use
\verb=persist(MEMORY_AND_DISK)= if your cluster does not have enough
total memory. It was shown in \SFpap that good performances can still
be obtained
in this case. Note that serializing the objects might also improve
performances (\eg \verb|level=MEMORY_ONLY_SER|) but it was not observed in our case.
In order to trigger caching, one must call an \texttt{action} as
counting the total number of galaxies which requires access to the full data:
\begin{python}
print(gal.cache.count())
\end{python}
\begin{verbatim}
5926764680
\end{verbatim}
In our case, putting all the data in cache by counting them takes
about 90 s, with about 20 s coming from the PZ computation.

\subsubsection{Getting some basic statistical information}
\label{sec:stat}

Some basic statistical informations may be obtained on some (or all) variables
with:
\begin{python}
gal.describe(['z','zrec']).show()
\end{python}
\begin{verbatim}
+-------+-------------------+------------------+                                
|summary|                  z|              zrec|
+-------+-------------------+------------------+
|  count|         5926764680|        5926764680|
|   mean|  0.875229444425171|0.8752293689731887|
| stddev|0.47360539092073933|0.4771461812884577|
|    min|        -5.93947E-4|       -0.12403674|
|    max|          2.4352543|          2.943411|
+-------+-------------------+------------------+
\end{verbatim}

If we only need some specific values (e.g \verb=minimum/maximum=), it is more efficient to use
the \spark functions: 
\begin{python}
minmax=gal.select(F.min("z"),F.max("z"))\
       .first()
zmin=minmax[0]
zmax=minmax[1]
\end{python}

\tab{perf} gives the measured user-time in each case: once the data
are in cache, one observes that all those commands run in seconds.

\subsubsection{Histograms}
\label{sec:histo}

We now wish to go further and study the redshift distribution of
galaxies.
Although RDDs provides some command to build histograms, we will show that
it is much more efficient to design a function using dataframes
capabilities, as follows
\begin{enumerate}
\item add a new column to the dataframe containing the bin number,
\item group the data by this number,
\item count the number of values in each group,
\item sort the bin index by ascending order.
\end{enumerate}

Adding the $z$ bin number column (labelled \q{bin}) is done most
efficiently using
standard column operations:
\begin{python}
Nbins=100
dz=(zmax-zmin)/Nbins 
zbin=gal.select(gal.z,\
     ((gal.z-zmin-dz/2)/dz).astype('int')\
     .alias('bin'))
\end{python}

Then, grouping by the \verb=bin= column, counting its members and sorting in
ascending order is performed by: 
\begin{python}
h=zbin.groupBy("bin")\
  .count()\
  .orderBy(F.asc("bin"))
\end{python}

Finally we may want to add the bin
locations, drop the bin number and go back to the \py
world by recovering a standard \texttt{pandas} object:
\begin{python}
pd=h.select("bin",\
   (zmin+dz/2+h.bin*dz).alias('zbin'),\
   "count")\
   .drop("bin")\
   .toPandas()
\end{python}

The histogram is obtained in about 10 s (\tab{perf}) which is
impressive for running on $6~10^9$ data records.

\begin{figure}
  \centering
  \includegraphics[width=.45\textwidth]{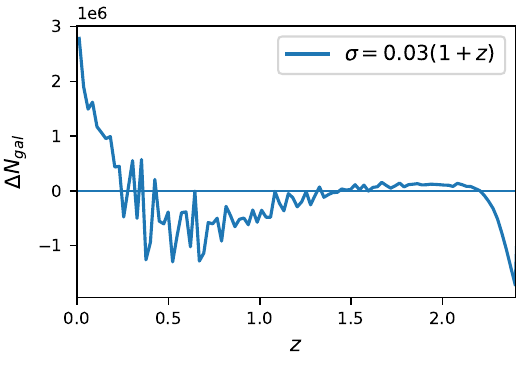}
  \caption{Difference of galactic density when applying a Gaussian
    photometric smearing.}
  \label{fig:dngal}
\end{figure}

We can now study for instance how the selection function varies with the
PZ smearing which is shown on Fig.\ref{fig:dngal}.

In order to prepare for the next part, let us see how to 
build the histogram by calling an external function. Operations may be
applied onto dataframes with a User Defined Function (UDF) as:
\begin{python}
binNum=F.udf(lambda z: int((z-zmin-dz/2)/dz))
zbin=gal.select(gal.z,\
     binNum(gal.z)\
     .alias('bin'))
\end{python}

But we find that there are performance issues since execution time goes from the previous 10 s to 2 minutes. 

To alleviate this issue, \spark introduced recently (\texttt{v2.3.0})
\texttt{pandas\_udf}'s.
Previously, standard UDF's were processing one row-at-a-time, which turns out to be slow for large numbers of rows.
Newly introduced \texttt{pandas\_udf}'s are built on top of Apache
Arrow\footnote{\url{https://arrow.apache.org/}} which, among several
other features, bring vectorized optimization.
Scalar \texttt{pandas\_udf}'s\footnote{There are two types of \texttt{pandas\_udf}'s: scalar and grouped map. In this paper we focus only on the former.} are used for vectorizing scalar operations using \texttt{pandas.Series} (groups of rows).
The UDF is re-written as: 
\begin{python}
import pandas as pd
from pyspark.sql.functions \
     import pandas_udf, PandasUDFType

@pandas_udf("float", PandasUDFType.SCALAR)
def binNumber(z):
    return pd.Series((z-zmin)/dz)

zbin=gal.select(gal.z,\
     binNumber("z").astype('int')\
     .alias('bin'))
\end{python}
The user-time becomes 40 s which is better although not optimal and
will be discussed more in \sect{sec:perf}. 
The main lesson from this part for \py users is to
always work with dataframes and  whenever possible with the native \texttt{Spark.SQL} functions.

\subsubsection{Tomography}
\label{sec:tomo}
Measuring galactic
over-density power-spectra over some redshift bins (called
tomographic \q{shells}) is a nearly
optimal method in cosmology to study galaxy clustering, especially
for photometric surveys \citep{Crocce:2011,Asorey:2012}. Measuring the
cross-correlation between nearby shells gives also access to
Redshift-Space-Distorsions even in photometric surveys where the
radial information is strongly suppressed
\citep{Ross:2011}. Cross-correlation between far-away bins is also
of interest: since, neglecting magnification lensing, 
no cosmological signal is expected there, any observed correlation
singles-out some remaining systematics (as PZ distribution tails).

\begin{figure}
  \centering
  \includegraphics[width=.5\textwidth]{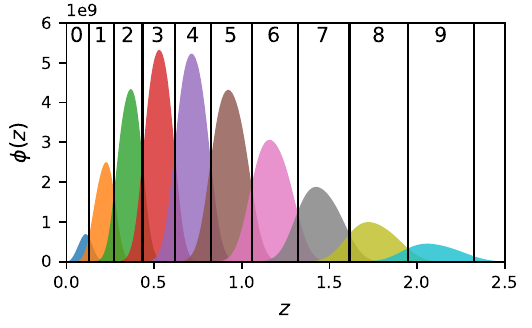}
  \caption{Position of the 10 tomographic bins in the observed space
    (vertical black lines with the numbering convention up) and
    contributions from the true redshifts  to each bin assuming a
    Gaussian PZ smearing with $\sigma_z=0.03(1+z)$ (filled
    curves).}
  \label{fig:tomo_bins}
\end{figure}

Such studies can be efficiently performed with \spark.
We have chosen 10 redshit bins marked out as vertical lines on
Fig.\ref{fig:tomo_bins}. Since the selection acts on the observed space,
each bin receives contribution from the true redshifts according to
the coloured distributions.

For each tomographic shell, we use the well-known \hp sphere
tessellation  \footnote{\url{https://healpix.jpl.nasa.gov/}}, which
partitions hierarchically the sphere (using the $nside$ parameter) with
equal-area and iso-latitude pixels. This latter property allows
an efficient computation of power-spectra over the sphere.
Using this scheme, one can map each (RA/DEC) position on the sky to
a single pixel number.
So, the construction of the projected galaxy number onto a \hp
map ($nside=512$) is as easy as building an histogram but using
the external \texttt{ang2pix} function to determine the pixel
number. For performances we use \texttt{pandas\_udf} to call this
function.

For a $z\in[z_1,z_2]$ shell
\begin{python}
import pandas as pd
import numpy as np
import healpy as hp
nside=512

#define the UDF
@pandas_udf('int', PandasUDFType.SCALAR)
def Ang2Pix(ra,dec):
    theta=np.radians(90-dec)
    phi=np.radians(ra)
    return pd.Series(hp.ang2pix(nside,theta,phi))           

#build the shell
shell=gal.filter(gal['zrec'].between(z1,z2))

#build the HEALPix map
map=shell.select(Ang2Pix("RA","Dec")\
    .alias("ipix"))\
    .groupBy("ipix")\
    .count()\
    .toPandas()

#back to python world
myMap = np.zeros(12 * nside**2)
myMap[map['ipix'].values]=map['count'].values
\end{python}

\begin{figure}[h!]
  \centering
  \includegraphics[width=.5\textwidth]{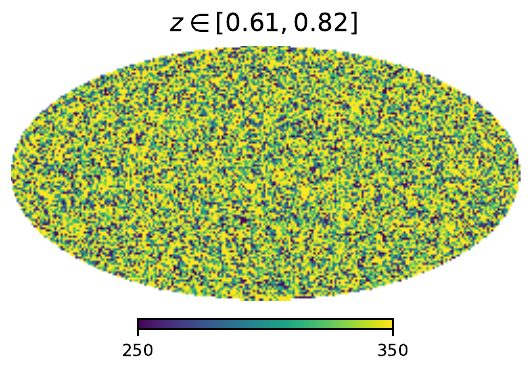}
  \caption{Mollweide projection of the contrast density map $\tfrac{N-\bar N}{\bar N}$ obtained
    from our LSST simulation on bin 4
    (see Fig. \ref{fig:tomo_bins}).}
  \label{fig:map4}
\end{figure}

We end up with a standard \hp map on
which one can perform further analysis. An example is shown on Fig. \ref{fig:map4}.
We note that the \texttt{python} packages must be available on each executor.

Concerning performances (\tab{perf}), each shell projection is obtained in about
30 s, quite independently of the galaxy population. All the 10 shells
are obtained in about 5 mins.

\begin{figure}[h!]
  \centering
%  \showthe\columnwidth % Use this to determine the width of the figure.
  \includegraphics[width=.5\textwidth]{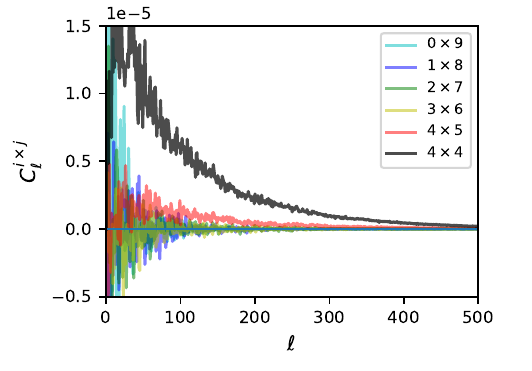}
  \caption{Tomographic power spectra reconstructed by cross-correlating
    the over-density maps with the bin numbering shown in the legend
    (see also Fig. \ref{fig:tomo_bins}). As discussed in \sect{sec:tomo} we see no power for
    well separated
    bins, some small contribution from the $4\times 5$ adjacent one and
    strong power for the $4\times 4$ auto-spectrum.}
  \label{fig:spectra}
\end{figure}

From the maps, one can then compute auto and cross spectra using
standard  \texttt{healpy} functions. We illustrate some
results we obtain for a few of them on Fig \ref{fig:spectra}.

\subsubsection{Performances}
\label{sec:perf}

We already discussed the user-time measured for each step. They are
summarized in \tab{perf}. 
But are we far from the best possible ones
in \spark?
The native \texttt{Spark} language is \sca which generally
leads to the best performances.
So, we have recoded and run all the previous commands in \sca
and compare performances in \tab{perf}.
\begin{table}
  \centering
  \begin{tabular}{|l|l|c|c|}
    \hline
   Section & analysis & \py & \sca \\
   \hline
   \multirow{2}{*}{\ref{sec:load}} & load(HDU)  & $2.8\pm0.1$ & $8.8\pm0.2$\\
   &PZ + show(5) & $12.4\pm0.6$ & $13.7\pm1.2$\\
   \hline
   \ref{sec:cache}&cache (count) & $97.7\pm4.0$ & $95.4\pm5.0$\\
   \hline
   \multirow{3}{*}{\ref{sec:stat}}& stat(z) & $3.9\pm1.5$ & $4.9\pm2.5$\\
   &stat(all) & $9.8\pm1.0$ & $11.0\pm0.9$\\
   &minmax(z) & $1.8\pm0.3$ & $3.2\pm0.7$\\
   \hline
   \multirow{3}{*}{\ref{sec:histo}}&histo (dataframe) & $11.5\pm1.5$ & $13.0\pm0.8$\\
   &histo (UDF) & $114.9\pm5.6$ & $13.9\pm1.2$\\
   &histo (pandas UDF) & $43.3\pm4.5$ & -\\
   \hline
   \multirow{2}{*}{\ref{sec:tomo}}&1 shell & $30\pm3$ & $13\pm2$\\
   &all shells (10) &$307\pm34$ & $130\pm18$\\
   \hline
  \end{tabular}
\caption{User-time (in seconds) for the various analysis steps described in
  the text using the \py commands
  (first column) and the \sca ones (second column). 
  Results were obtained by running sequentially the commands
  from the top to the bottom of the table, 10 times and were then averaged.
}
\label{tab:perf}
\end{table}

The initialization phase (\verb=load(HDU)=) is slightly longer in \sca
than in \py. Caching the $6~10^9$ data records which is the most
demanding part but only needs to be performed once,
is obtained in our case in about 1.5 mins in each case.
Then, the statistics part is slightly more
efficient in \py. As long as native dataframe operations
are used both implementations yield a similar 10 s value for
histograms. The use of an external (UDF)
function is severely penalized in \py. Using
\texttt{pandas\_udf}'s the user-time is reduced to 40 s which becomes
reasonable although a factor of 3 higher than the corresponding \sca
implementation. The same kind of factor is observed when building
the shells. Still, reconstructing a tomographic map in about
30 s remains very satisfactory for interactive work. 

Let us compare these performances to the ones we would have obtained
in a more standard  imperative way. For instance, let us
consider how we would have computed the \verb=min/max= values of the PZ
redshifts:
\begin{enumerate}
\item each FITS files is opened and read,
\item redshift uncertainties are added shooting random numbers,
\item the min and max values are computed and stored,
\item all min/max values are combined to obtain the lowest (highest) min (max).
\end{enumerate}
We implemented that process in \py: it takes about 45
mins to process the 32 files.  Which is to be compared to the $2.8+12.4+97.7+1.8\simeq 2$ mins
we obtained with \spark. This sequential comparison is unfair since one could also 
implement some MPI logic to read the files in parallel on each node,
or even post-process the files to distribute them on each core. The
point is not getting the very best performance but already an
excellent one with very little effort.
Furthermore, while in the imperative approach all the data in memory
is lost when the program ends, using the cache in \spark offers the possibility
to investigate several aspects of the data in a single session without
ever reloading them.

\section{Conclusion}

In the current days of ever-growing astronomical data, it is worth
investing in a technology which gives physicists access to the
\textit{interactive} analysis of billions of objects, boosting
time-to-physics considerably.

We have illustrated in this pedagogical introduction to users, 
how a 6 billion-objects realistic galactic catalogue could
be analysed interactively producing histograms in about 10 s
and tomographic bins in 30 s. 
Our main point is about \textit{simplicity}. The access to a large
volume of data in a distributed environment requires today quite involved technical skills that
are relieved by simply using \spark (\py) functions.

We emphasize that \spark is a \textit{Big Data} technology, meaning it
makes no sense using it for data volumes below a few tens of Gigabytes
as one would lose its main computational advantages by paying its
overhead costs. 
Although it is convenient to develop the
software on a personal computer do not expect interesting performances there.
The \spark technology is intended to be run
on data-centres where the storage is a crucial part.
However one does not necessarily need a very large cluster or a super-computer
for that; we achieved very satistactory results on 8 machines.
We focused here on rather simple operations without ever
worrying where data reside on the cluster nor how they are
organized, a process known as data partitioning. 
For performance, one always needs to minimize network connections among
the workers and this may need to be investigated in some cases, 
as for instance when cross-matching catalogs
\citep{zecevic:2018,BRAHEM::8481506}. It can be however a heavy
process since data needs to be 
shuffled among the different executors, and possibly re-written to
the disk. It further depends on the type of query that
is to be optimized. Spatial locality may be achieved straightforwardly
in 2D with a \texttt{Healpix} indexing and in 3D with an octree
\footnote{\url{https://astrolabsoftware.github.io/spark3D} offers a
  simple dataframe interface for doing it}. 
But some more evolved indexing may be required for instance when
also considering the color or shape of galaxies.
This is an important topic, but one that goes beyond the scope of this
paper and will be discussed in a forthcoming one.

We have addressed the question of interfacing \texttt{pyspark} to some external \py code,
which can be done reasonably well using (pandas) user-defined-functions.
This however requires I/O's to be quite simple. A difficulty we
encountered was that building some application always requires
access to some external \py libraries that must be available on each
executor. While \spark offers some integrated mechanism to transmit binary
modules to all workers, it is not a very efficient solution. The
administrator may install some set of common scientific libraries on each
node, but it is not a very flexible solution.
An elegant solution we found for NERSC was to build a dedicated \texttt{shifter}
image (a \texttt{docker} variant) including \spark and the necessary
libraries.
This is a simple solution that can be put in place by each user, but running a
docker-like image over each node is not always supported on every data
centre.

One can also interface some \texttt{C}, \texttt{C++} or
\texttt{Fortran} codes with 
\sca and the \texttt{JNA} (Java Native Access) library. Some examples
were 
reported in \citet{Zhang:2015,Wiley:2011} and the interested reader
may start with \url{https://github.com/astrolabsoftware/Interfaces} for a
practical usage.
On super-computers one may take benefit from combining the HTC and HPC
approaches, a road that has still to be explored.
This is an exciting perspective, that would
allow to add science to a technology born once in a public lab. 
It is the goal of the \texttt{Astrolab}
\footnote{\url{https://astrolabsoftware.github.io}} organization to
have everyone interested in such a project join it.

\section*{Acknowledgements}
We acknowledge the use of the \hp package  \citep{2005ApJ...622..759G} and the 
\colore fast simulation with support
from David Alonso, Anze Slosar and Javier Sanchez that we kindly
thank. The \spark work
was performed at the VirtualData center at Universit\'e Paris Sud
and we thank Adrien Ramparison for the upgrades and maintenance of the
cluster. The \colore 
simulation was run at the National Energy Research Scientific
Computing Center, a DOE Office of Science User Facility supported by
the Office of Science of the U.S. Department of Energy under Contract
No. DE-AC02-05CH11231.

\appendix
\section{Handling realistic photo-z distribution}
\label{sec:appendix}

We have seen in Sect \ref{sec:PZ} how to implement some very simple
Gaussian smearing of the redshift variable. Beyond this simple approximation,
the more complex situation is the following. In brief, a photometric instrument like LSST
measures integrated fluxes in a few frequency bands  (6 for LSST) and from these
few values one infers the redshift. This is performed with the help of some external catalogues where the
spectral energy distributions (SED) of a sample of galaxies has been
precisely measured, for instance with a spectrometer. Since the effect
of redshift is to shift the SED, the idea is to determine the redshift (noted  $z_{rec}$) by measuring the displacement
of the observations to match the catalogue SED. This can be performed with template-fitting or
Machine Learning algorithms (see for instance \cite{2018NatAs.tmp...68S,2019A&A...621A..26P}). 
But, some confusion may arise when matching
the catalogue which results in tails in the inferred redshift values.
Knowing precisely the photometric redshift distribution is an
important ingredient to several key cosmological analyses as those related
to weak-lensing \citep{Mandelbaum:2017}.

\begin{figure}
  \centering
  \includegraphics[width=\linewidth]{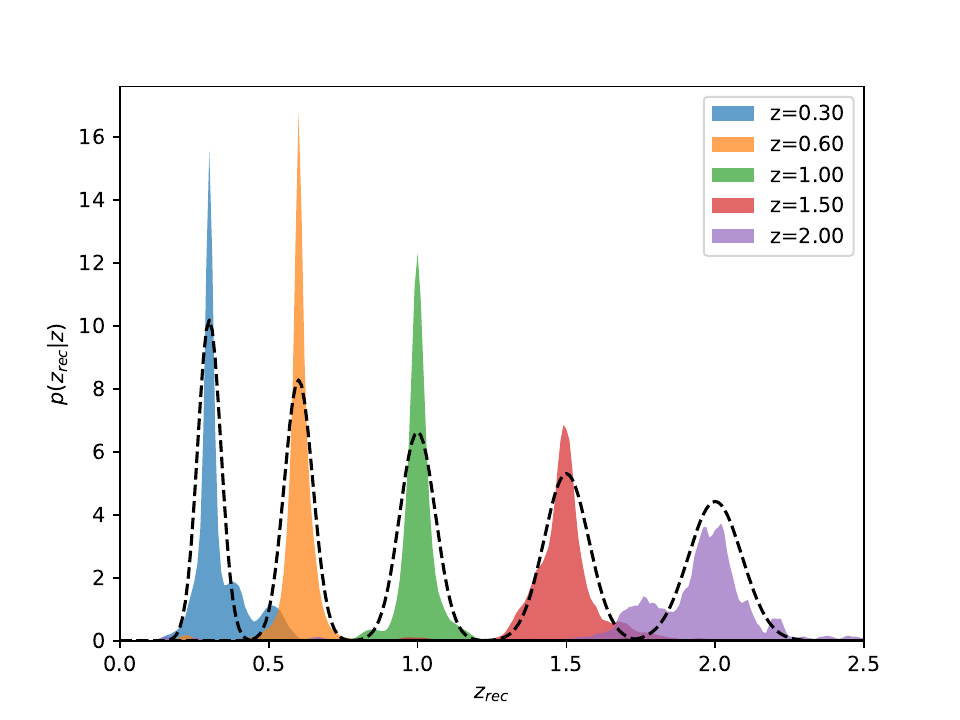}
  \caption{\label{fig:PZ_distrib} Distributions of some photometric
    redshift distributions given the true values shown in the box, as obtained in
  \citet{Renault:2019} using a full template-fitting method. Note that
  we do not use a quality cut. The dashed lines corresponds to the
  simplified Gaussian model with $\sigma_z=0.03(1+z)$.}
\end{figure}

In order to assess a more realistic LSST use-case
and show how to implement it within \spark, we use some recent results 
from \citet{Renault:2019}. The authors have reconstructed the distributions of the
observed photometric redshifts given the true ones $P(z_{rec}|z)$
applying a full template-fitting procedure.
Fig. \ref{fig:PZ_distrib} shows some of these distributions which
reveals that their main part is more packed than the Gaussian model
but exhibits complex tails.

Our aim is, for our simulation, to shoot some random number
according to these distributions. We use for that a standard
Monte Carlo technique: to generate samples according to some
target distribution $P$, one generates some uniformly distributed
samples and transform them according to the inverse of the $P$
cumulative distribution, or, in statistical language
$u\sim U(0,1), F^{-1}(u)\sim p$ when $F(z)=\int_{-\infty}^z P(x)dx$.
The cumulative function can be computed and inverted numerically.

In our case, the authors of \citet{Renault:2019} have provided a table of 300 photometric
distributions according to their true redshifts in the $z \in
[0,3]$ range. For each true value
$z_i$ we computed the cumulative sum of the corresponding photometric
distribution $P(z_{rec}|z_i)$ and numerically invert it. Since in the
inversion one looses the abscissa regularity, we re-sampled the values
onto a linear grid of 1000 points over the [0,1] interval.
We then stored the resulting values in a table $300 \times 1000$ as a plain text file.
Then, for each true redshift value $z_i$ and an associated uniformly
generated number $u_i$, one just need to read the proper value from
this table. We have checked that the full procedure applied to the gaussian case
reproduces the results shown in Sect \ref{sec:histo}.

The implementation of the method  within \spark is now rather simple: 
\begin{enumerate}
\item read the inverse-cumulative table
\item add to the dataframe a column of uniformly generated numbers ("u").
\item for each true redshift ("z") and random sample ("u"), locate the index
  within the table and get the value from a user-defined-function (\texttt{pandas\_udf}).
\end{enumerate}

The code reads as follows 
\begin{python}
import numpy as np
import pandas as pd

# read the inverse-cumulative file 
cuminv=np.loadtxt('cum_inv.txt')
# we know the binning that were used
dz=0.01
du=1/1000.

#find indices and return the table values
@pandas_udf('float', PandasUDFType.SCALAR)
def z_phot(zr,u):
    iz=np.array(zr/dz,dtype='int')
    iu=np.array(u/du,dtype='int') 
    return pd.Series(cuminv[iz,iu])

#add column of uniform random numbers
gal=gal.withColumn("u",F.rand())

#transform with the inverse-cumulative table
gal=gal.withColumn("zrec",z_phot("z","u")+dz/2)\
       .drop("u")  #do not need u anymore
\end{python}

\begin{figure}
  \centering
   \includegraphics[width=\linewidth]{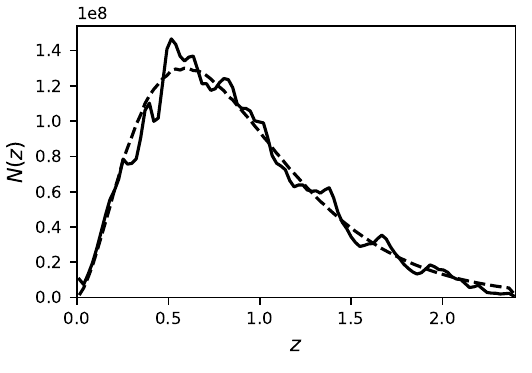}
  \caption{\label{fig:N_PZ}Histogram of the reconstructed redshift of galaxies in our
    10-years LSST sample (full line) compared to the underlying real
    one (dashed), using the realistic photometric
  distributions from \cite{Renault:2019}.}
\end{figure}

Using the method discussed in Sect. \ref{sec:histo}, we histogram the
"zrec" column which gives the \textit{selection function} shown on Fig. \ref{fig:N_PZ}.
The non-Gaussian behaviour of the photometric distribution
dramatically affects the selection function in a non-trivial way, 
which to our knowledge is shown for the first time.

\pagebreak

\bibliographystyle{elsarticle-harv} 
\bibliography{Spark4Phys}

\end{document}